\documentclass[a4paper,10pt]{article}
\usepackage[margin=1in,letterpaper]{geometry}

\usepackage[utf8]{inputenc}
\pdfoutput=1

\usepackage{graphicx}
\usepackage[hypcap]{caption}
\usepackage[section]{placeins}
\usepackage{hyperref}
\usepackage{natbib}
\usepackage{setspace}
\usepackage{lineno}

\title{How robust are Structural Equation Models to model misspecification? A simulation study}
\author{Lionel R. Hertzog}
\date{Terrestrial Ecology Lab, University of Ghent, K.L. Ledeganckstraat 35, BE-9000 Gent. e-mail: \url{lionel.hertzog@ugent.be}}

\graphicspath{{./Figures/}}

\begin{document}
\maketitle

\hrulefill

\doublespacing

\section*{Abstract}

\begin{enumerate}
 \item Structural Equation Models (SEMs) are routinely used in the analysis of empirical data by researchers from different scientific fields such as psychologists or economists. In some fields, such as in ecology, SEMs have only started recently to attract attention and thanks to dedicated software packages the use of SEMs has steadily increased. Yet, common analysis practices in such fields that might be transposed from other statistical techniques such as model acceptance or rejection based on p-value screening might be poorly fitted for SEMs especially when these models are used to confirm or reject hypotheses.
 \item In this simulation study, SEMs were fitted via two commonly used R packages: lavaan and piecewiseSEM. Five different data-generation scenarios were explored: (i) random, (ii) exact, (iii) shuffled, (iv) underspecified and (v) overspecified. In addition, sample size and model complexity were also varied to explore their impact on various global and local model fitness indices.
 \item The results showed that not one single model index should be used to decide on model fitness but rather a combination of different model fitness indices is needed. The global chi-square test for lavaan or the Fisher’s C statistic for piecewiseSEM were, in isolation, poor indicators of model fitness. In addition, the simulations showed that to achieve sufficient power to detect individual effects, adequate sample sizes are required. Finally, BIC showed good capacity to select models closer to the truth especially for more complex models.
 \item I provide, based on these results, a flowchart indicating how information from different metrics may be combined to reveal model strength and weaknesses. Researchers in scientific fields with little experience in SEMs, such as in ecology, should consider and accept these limitations.
\end{enumerate}

\vspace*{0.5cm}

Keywords: R, lavaan, piecewiseSEM

\hrulefill

\section*{Introduction}

Structural equation models (SEMs) allow researchers to explicitly model measurement processes and complex inter-relations between variables \citep{kline1999}. While classical regression analysis require to set one response variables and do not model correlations between explanatory variables, SEMs allow more flexibility where variables can be both depending on and affecting other variables. Other important features of SEMs include (but are not limited to): (i) the possibility to model latent variables, theoretical and unobserved constructs such as motivation or stability, that affect a set of observed indicator variables, (ii) the inclusion of mediation effects which allows the quantification of direct, indirect and total effects of one variable on another. Another appealing aspect of SEMs is the possibility to visually represent variables or constructs with boxes and draw complex relationships between them with arrows which potentially provide tighter links to theories \citep{grace2006}. Indeed, some have claimed that SEMs, if built correctly and carefully, have causal interpretation \citep{pearl2012} which would provide the possibility to test mechanisms and theories from observational data \citep{shipley2000b} something that is, at best, arduous in regression analysis. 

While SEMs originated in the social sciences and have a long history both of development and of use by empirical researchers \citep[i.e.][]{bentler1980}, in ecology SEMs are still perceived as a new method, despite the presence of early ecological studies using SEMs \citep[i.e.][]{power1972}. There has been a recent surge of ecological studies using this approach driven by some influential papers \citep{grace2016}, books written by and for ecologists \citep{shipley2000b,grace2006} and the development of open-source, freely available R packages \citep{lefcheck2016, rosseel2012}. It is interesting to note that while psychologists are wary in their use of causality statement in SEMs preferring to leave out the exploration of mechanisms (Y. Rosseel pers. communication), ecologists are attracted to SEMs in part for its capacities to unravel mechanisms \citep{eisenhauer2015,fan2016} a preeminent goal in current ecological research. The correctness of these assumptions have been discussed elsewhere \citep{petraitis1996,pearl2012}, and will not be considered further in the present article. Instead, I focus here on the impact of model misspecification such as underfitting (missing relation between variables) or overfitting (too many relations between variables) to check if commonly used model fitness indices are able to detect these misspecifications and provide appropriate guidance to improve the models.

As in any other statistical framework, structural equation models require checking that the model fitted the data sufficiently well to proceed with interpretation of the fitted effects. Ample (psychological) literature exist on the issues of structural equation model testing and model evaluation \citep[e.g.][]{bollen1992} with continuous development \citep{bollen2014,lin2017}. However, differences in terminology and publication in journals not widely read by researchers in other fields of research prevents the dissemination of methodological knowledge and best practices of SEMs testing across fields. One of the current trend in SEM evaluation is the use of information-theoretic criterion (IC) metrics, such as the Akaike Information Criterion (AIC), Bayes-Schwarz Information Criterion (BIC) or more complex derivations such as Haughton-BIC \citep{lin2017}. Simulation studies found that IC metrics tended to perform better than other fitness indices based on the classical chi-square score but also tended to favor overly complex models over underspecified models \citep{bollen2014}. SEMs being relatively new in the field of ecology there is little to no formal training on SEMs for (under)graduate students in ecology. As a result, techniques and traditions from other statistical framework such as ANOVAs or regressions might be applied to SEMs leading to sub-optimal or even biased inference. Books on SEMs written for ecologists do provide some guidance regarding model testing, for instance \citet{shipley2000b} discuss at length the chi-square test on fitted covariance matrices and alternative fitness tests, while \citet{grace2006} provide some guidance on different model evaluation strategies based on different approaches from strictly confirmatory to purely exploratory. Exploring the limits and the strengths of a modelling framework can be performed with closed-form equations for simple models, such as for instance the power of ANOVA to detect differences between means. For more complex modelling framework such as SEM, simulation approaches are the only way to explore sensitivity of model checks or power to detect effects \citep{bolker2008}.

Several software packages already exist to simulate datasets from SEMs. The simsem package for R \citep{simsem}, for instance, allows great flexibility in fitting different types of SEMs with or without latent variables and structural equations. These packages usually assume that the correct model structure is known and so generate synthetic data from that model. In this paper I present an approach to simulate simple SEMs with the possibility to generate data following different scenarios in order to compare limits of model checking under different model misspecification. In addition, two approaches to fit SEMs are compared, one with global estimation of parameters using the covariance matrix \citep{rosseel2012} and one with piecewise estimation of model parameters in individual linear models combined by checking independence claims \citep{lefcheck2016}. The aim is to compare: (i) model fitting strategy (global vs piecewise), (ii) model misspecification type, (iii) sample size and (iv) model complexity on various common metrics used to check model fitness. 

\section*{Methods}

Two main simulation batch were run to explore different SEMs fitness metrics. The first simulation batch aimed at exploring the effect of various model miss-specification scenarios on classical SEMs fitness metrics. The second batch of simulations aimed at exploring how various information criteria (IC) metrics performed under various model miss-specification. 

\begin{figure}
 \includegraphics[width=\textwidth,keepaspectratio]{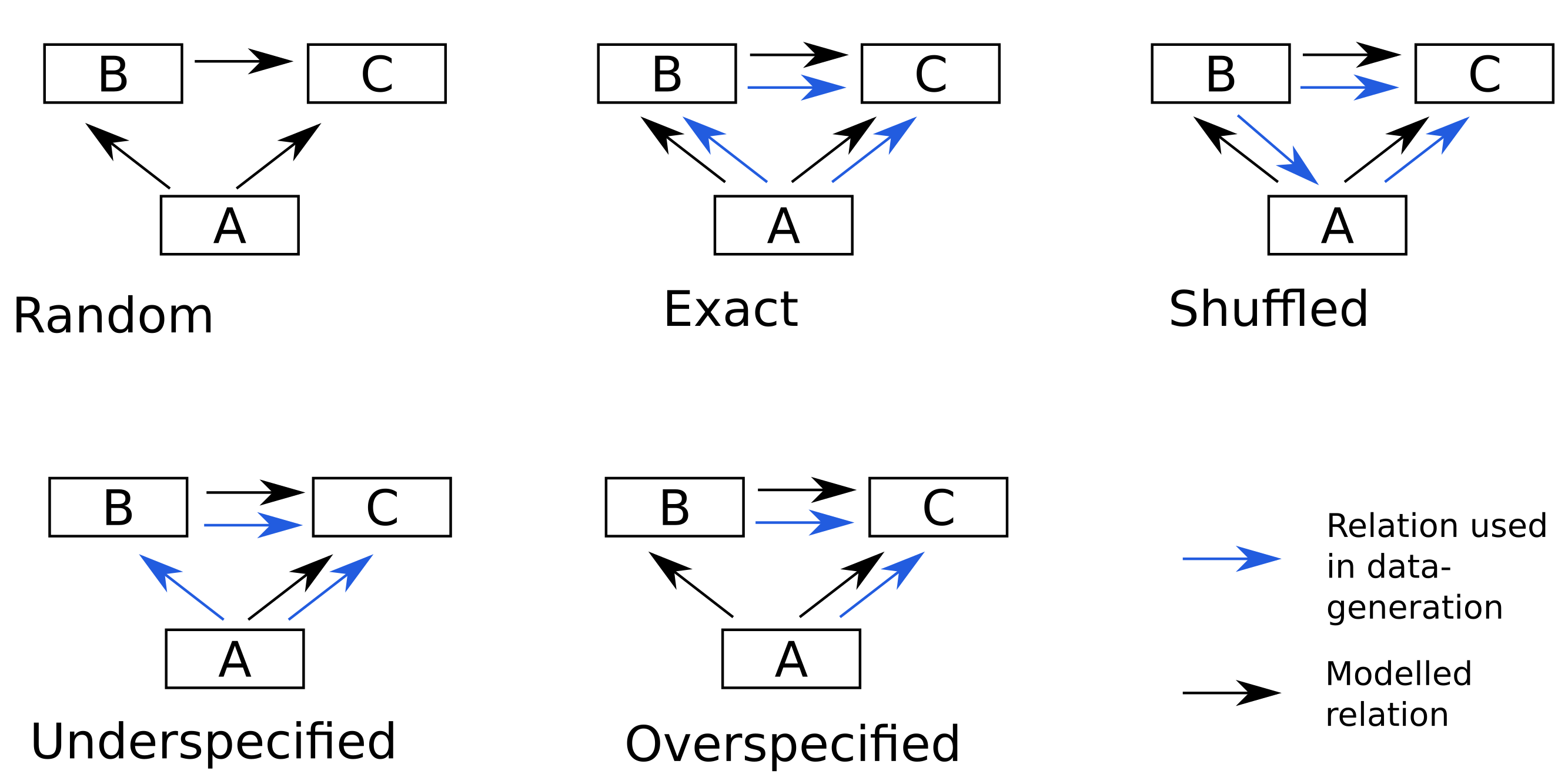}
 \caption{Schematic representation of the different data-generation scenarios used in the first batch of simulations. A, B and C represent hypothetical variables, the black arrows show the relationships assume in the fitted models and the blue arrow the relationship present during the data-generation process.}
 \label{fig:concept}
\end{figure}

In the first batch of simulations, data were generated based on 5 scenarios: (i) random, (ii) exact, (iii) shuffled, (iv) overspecified and (v) underspecified, see Figure~\ref{fig:concept}. In the random scenario, data were generated as random normal deviates with a mean of 0 and a standard deviation of 1 without any signal between covariates. In the exact scenario, covariates were sequentially generated following exactly the created model. The created models ensured that there was at least one exogenous covariate to ensure model identifiability. The exact data generation first generated random uniform deviates for the exogenous covariates. Then for each subsequent covariate, regression coefficients were generated following a normal distribution with a mean of 0 and a standard deviation of 2.5. Linear predictors were derived by combining the drawn regression coefficient and the effect of the other covariates. Then normally-distributed residuals with a mean of 0 and a standard deviation of 1 were added to the linear predictors. In the shuffled scenario, a fixed proportion of relations were reversed, in other words when the model set A -\textgreater~B, a shuffled relation is B -\textgreater~A. 25\% of the relations were shuffled across model sizes, relations to be shuffled were randomly selected. Then the shuffled relation matrix was used to generate the data following the same steps as in the exact data generation. In the overspecified scenario, 25\% of the relations from the model were randomly selected and dropped during the data generation process. For instance, the model might assume A + B -\textgreater~C but in the data generation we have A -\textgreater~C. In other words in the overspecified scenario the assumed model is too complex compared to the generated data. The underspecified scenario is the opposite of the overspecified one, namely 25\% additional relations are added during the data generation process which result in a model being too simple compared to the data. 

Four readily available metrics were derived from the models: (i) the acceptance of the model based on the p-value of the Fischer's C score in piecewiseSEM and on the p-value of the chi-square test in lavaan. Models with p-values higher than 0.05 were considered as accepted. In lavaan some of the models did not converge, these were considered as being rejected. (ii) The proportion of significant regression coefficients (p-values \textless~0.05). (iii) The average R-square values of the model derived from the adjusted R-square of the individual linear models composing the model for piecewiseSEM and from the average R-square values of the endogenous variables in lavaan. Just like the paths between covariates constitute testable hypothesis, so does conditional independence this was quantified as (iv) the proportion of conditional independence tests implied by the model structure that failed (p-value \textgreater 0.05). Conditional independence is the fact that two variables are independent given that potential mediators are fixed. Conditional independence was computed based on the localTests function in the dagitty package v0.2 \citep{dagitty}. 

In the second batch of simulation, the data were generated once based on the exact scenario defined above. These data were then fitted to four different models: (i) exact, this model exactly followed the data generation process, (ii) shuffled, some relations between covariates were shuffled, (iii) overspecified, extra relation were added and (iv) underspecified, some relations were dropped. These models were created based on the data generation scenarios used in the first batch of simulations. The four different models were fitted to the data and three IC metrics were explored: (i) The sample size corrected AIC values of the model (AICc), (ii) The BIC value of the model and (iii) The Haughton-BIC (HBIC), a variant of BIC computed as followed (after \citet{lin2017}):

\begin{equation}
 HBIC = logl - K * log(\frac{N}{2 * \pi})
\end{equation}

where \textit{logl} is the summed log-likelihood of the model, \textit{K} is the number of parameters in the model and \textit{N} is the sample size. If the scenario with the lowest IC metric was more than 2 units lower than the second lowest scenario, this scenario was labeled as "best scenario", if that was not the case then no scenario were labeled. 

In the two batch of simulations, the sample size took the values of 20, 40, 60, 80, 100, 200, 500, 1000, 5000 and 10000, which represent an extended gradient of classical sample size in empirical datasets in ecology. The number of covariates ranged from 5 to 10 also corresponding to typical model complexity in ecology. All covariates were observed, in other words there were no latent or composite variables in the models. SEMs can be seen as directed networks with flow of information going specific ways between the covariates. With this analogy the connectance of an SEM is the number of path divided by the total number of potential path. The relations between the covariates were randomly generated with a fixed connectance of 0.3, this resulted in models with between 7 and 30 paths among the covariates. The models were then fitted to the data using piecewiseSEM v1.2 \citep{lefcheck2016} or lavaan v0.5 \citep{rosseel2012}. To further explore the impact of the signal / noise ratio on the two batch of simulations the residual standard deviation (noise) was varied taking values of 0.5, 1 and 2.5 crossed with variations in the standard deviation of the normal distribution generating the path coefficients (signal) which took values of 1, 2.5 and 5. All results from these additional simulations are presented in the Appendix. In total this led to 2340 parameter sets for the first batch of simulations: 10 sample size * 3 covariate number * 5 data generation * 2 model fitting * 3 signal strength * 3 noise strength (only keeping 1 signal strength for the random scenario) and to 540 parameter sets for the second batch simulations: 10 sample size * 3 covariate number * 2 model fitting * 3 signal strength * 3 noise strength. Each parameter set was replicated 100 times to account for stochasticity in the data generation process. Per parameter set in the first simulation batch the results were summarized as followed: (i) proportion of accepted models, (ii) proportion of significant paths, (iii) average of R-square values, (iv) average of proportion of failed conditional independence tests. In the second simulation batch the results were summarized as the proportion of times that each scenario was labelled "best scenario". All the code used here is deposited in an online repository \citep{data} and the simulation data are deposited at the following url: \url{https://doi.org/10.6084/m9.figshare.7553558.v1}.

\section*{Results}

\subsection*{Proportion of accepted models}

\begin{figure}
 \includegraphics[width=\textwidth,keepaspectratio]{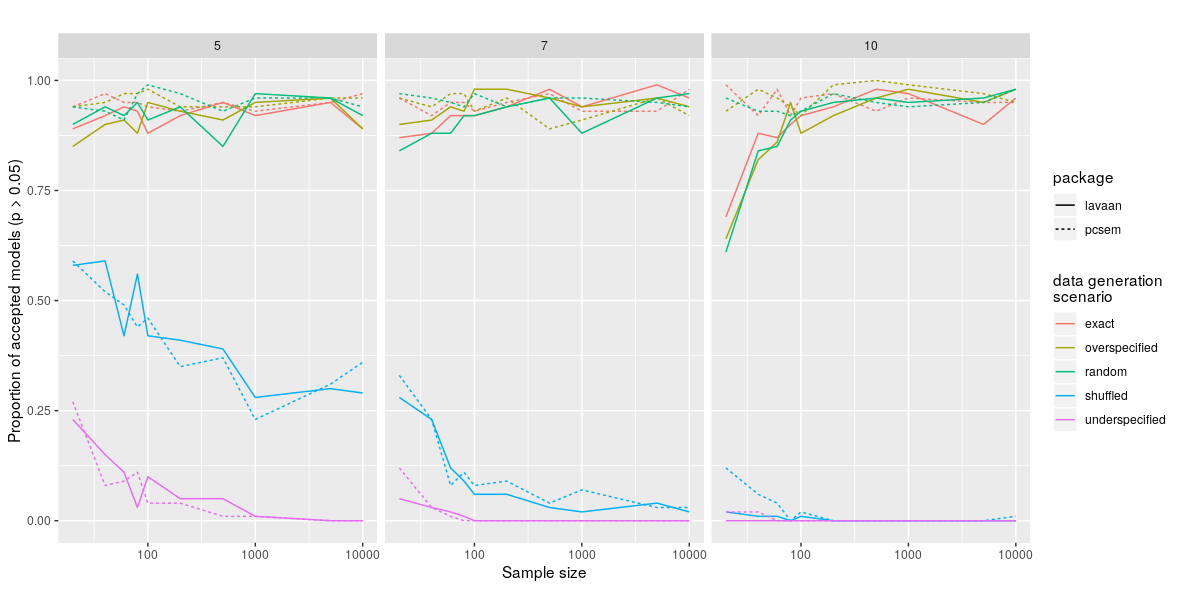}
 \caption{Effect of sample size, model complexity (number of covariates 5, 7 or 10 in the different panels), data generation scenario and R package (lavaan or piecewiseSEM) on the proportion of accepted structural equation models. Reported is the proportion of accepted models (p-values \textgreater~0.05) across 100 replications per parameter set (sample size, complexity, data-generation, model type). Models which failed to converge were considered as being rejected.}
 \label{fig:acc}
\end{figure}

Two groups of data-generation scenarios were apparent when comparing the proportion of accepted models (Figure \ref{fig:acc}). The first group had acceptance rates around 90\% and contained the random, the exact and the overspecified data generation scenarios. In this group the sample size had little effect on the acceptance rates to the exception of the models with 10 covariates fitted with lavaan where the acceptance rate asymptotically increased with sample size. In the second group, containing the shuffled and the underspecified data generation scenario, the acceptance rate was much lower and was depending on the sample size and on the number of covariates but not on the method used to fit the models (lavaan vs picewiseSEM). For the shuffled scenario with 5 covariates the acceptance rate was halved from 50 to 25\% when sample size increased from 20 to 10'000. For the underspecified scenario the acceptance rate dropped from 25 to 0\% over the same gradient. In models with 10 covariates, the acceptance rate for these two scenarios was constantly close to 0\%.   

\subsection*{Proportion of significant paths}

\begin{figure}
 \includegraphics[width=\textwidth,keepaspectratio]{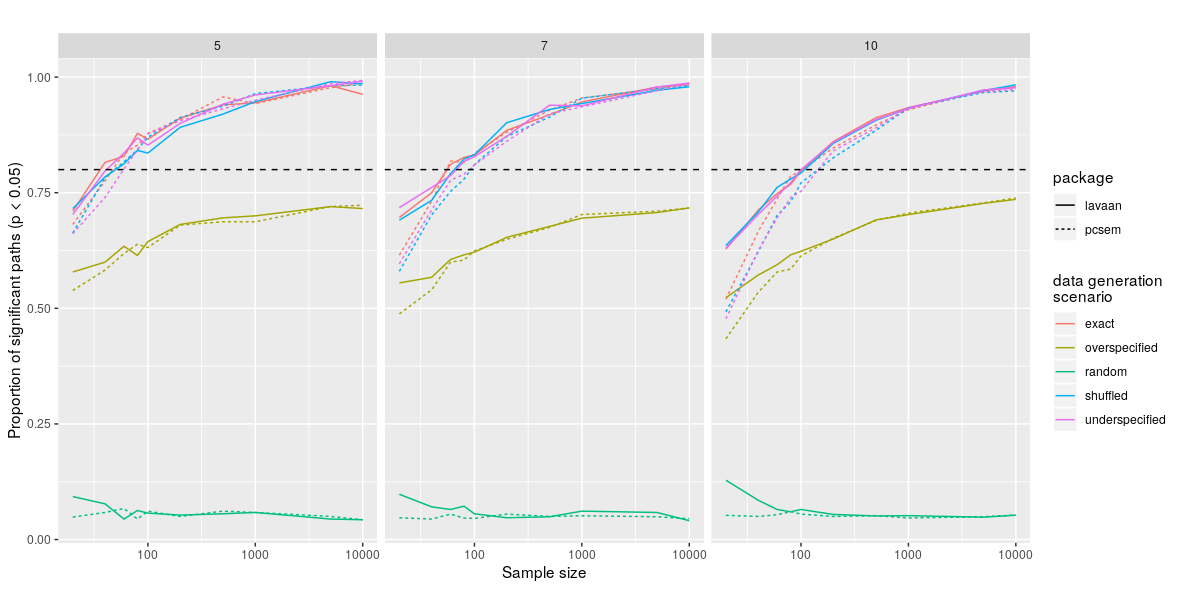}
 \caption{Effect of sample size, model complexity (number of covariates and parameters), data generation scenario and R package (lavaan or piecewiseSEM) on the proportion of significant paths (p-value of individual paths \textless~0.05). Reported is the average proportion of significant paths across 100 replications per parameter set. The dotted black line show the nominal power threshold of 80\%.}
 \label{fig:path}
\end{figure}

Lavaan and piecewiseSEM led to an almost identical pattern for the proportion of significant paths despite some indication that lavaan had slightly higher power at lower sample size and for models with larger number of covariates (Figure~\ref{fig:path}). The proportion of significant paths asymptotically increased with sample size for all data generation scenario except for the random one. Exploring when the proportion of significant paths on the exact scenario crossed the 80\% level commonly used to ensure sufficient power, revealed that the rule of thumb of 5 data points per parameters was roughly supported by the simulation data. The shuffled and underspecified data generation scenarios showed similar proportion of significant paths as the exact scenario. While the overspecified scenario had lower values for the proportion of significant paths never crossing the 80\% threshold. Finally, the random data generation scenario had low proportion of significant paths independent from the sample size at around 5\%.

\subsection*{R-square}

\begin{figure}
 \includegraphics[width=\textwidth,keepaspectratio]{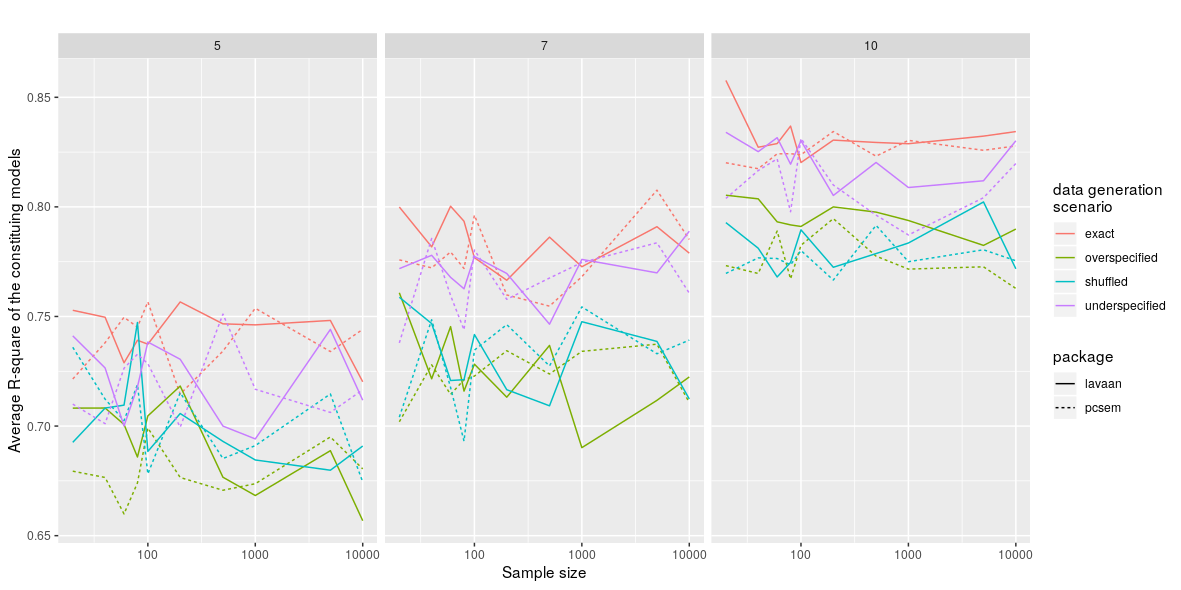}
 \caption{Effect of sample size, model complexity (number of covariates and parameters), data generation scenario and R package (lavaan or piecewiseSEM) on the average R-square of the individual model regression. For each fitted model the adjusted R-square of the individual linear regression was extracted and averaged across 100 replications per parameter set.}
 \label{fig:rsq}
\end{figure}

No consistent differences could be found between models fitted via lavaan or piecewiseSEM on the average adjusted R-square of the models  (Figure~\ref{fig:rsq}). Sample size had no detectable effects on R-square. The random data-generation scenario had very low adjusted R-square, close to 0\%. The other scenarios had similar R-square values with the exact data generation scenario showing a tendency for having larger R-square than the other ones. R-squares tended to be larger in models with higher number of covariates, for instance for the exact data-generation scenario, R-square was around 70\% for 5 covariates, but it was around 80\% for 10 covariates. 

\subsection*{Conditional independence}

\begin{figure}
 \includegraphics[width=\textwidth,keepaspectratio]{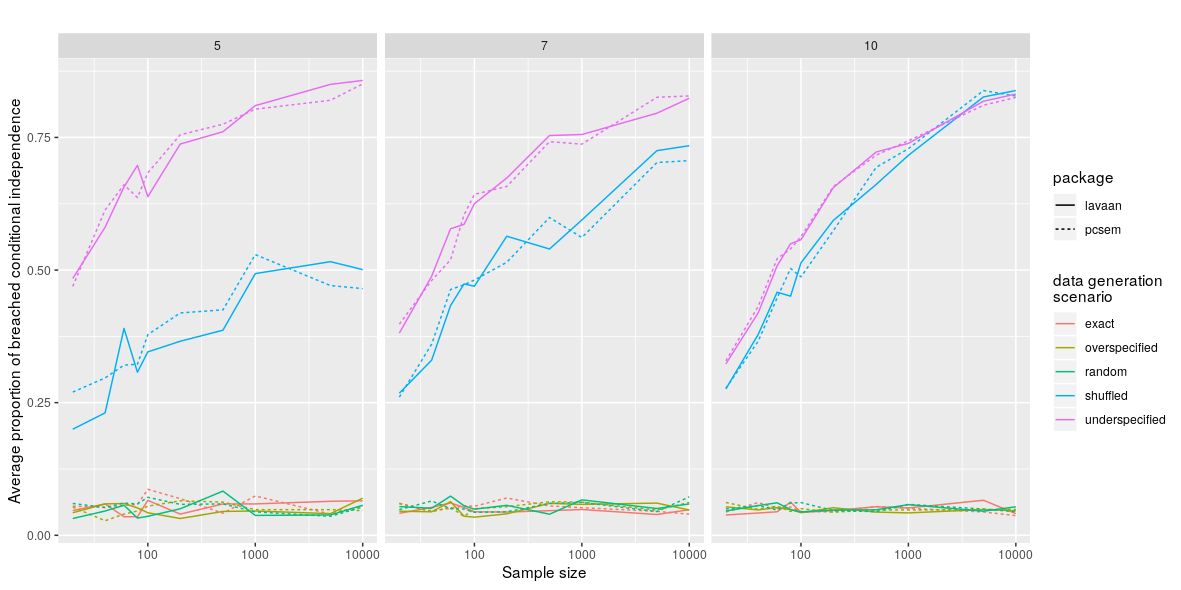}
 \caption{Effect of sample size, model complexity (number of covariates and parameters), data generation scenario and R package on the average proportion of conditional independence tests that failed (p-value \textless~0.05).}
 \label{fig:cond}
\end{figure}

Conditional independence in SEM imply that two variables are independent after controlling for the effect of one or several mediators. Failure to meet conditional independence indicate misspecification in the model structure.
Again no consistent differences could be found between models fitted via lavaan compared to models fitted via picewiseSEM. Two groups of scenarios appeared for the proportion of failed conditional independence, namely: (i) exact, random and overspecified, and (ii) shuffled and underspecified (Figure~\ref{fig:cond}). In the first group the average proportion of failed conditional independence tests was consistently low at around 5\% close to the type I error rate. For the other two scenarios, failure tended to increase with sample size. The effect of the number of covariates had opposite direction between the shuffled and underspecified scenario. While increasing the number of covariates increased the proportion of failures from around 40\% to around 60\% for the shuffled data-generation scenario, it reduced the proportion of failures from around 75\% to around 65\% for the shuffled data-generation one. 

\subsection*{AICc, BIC and HBIC}

\begin{figure}
   \includegraphics[width=\textwidth,keepaspectratio]{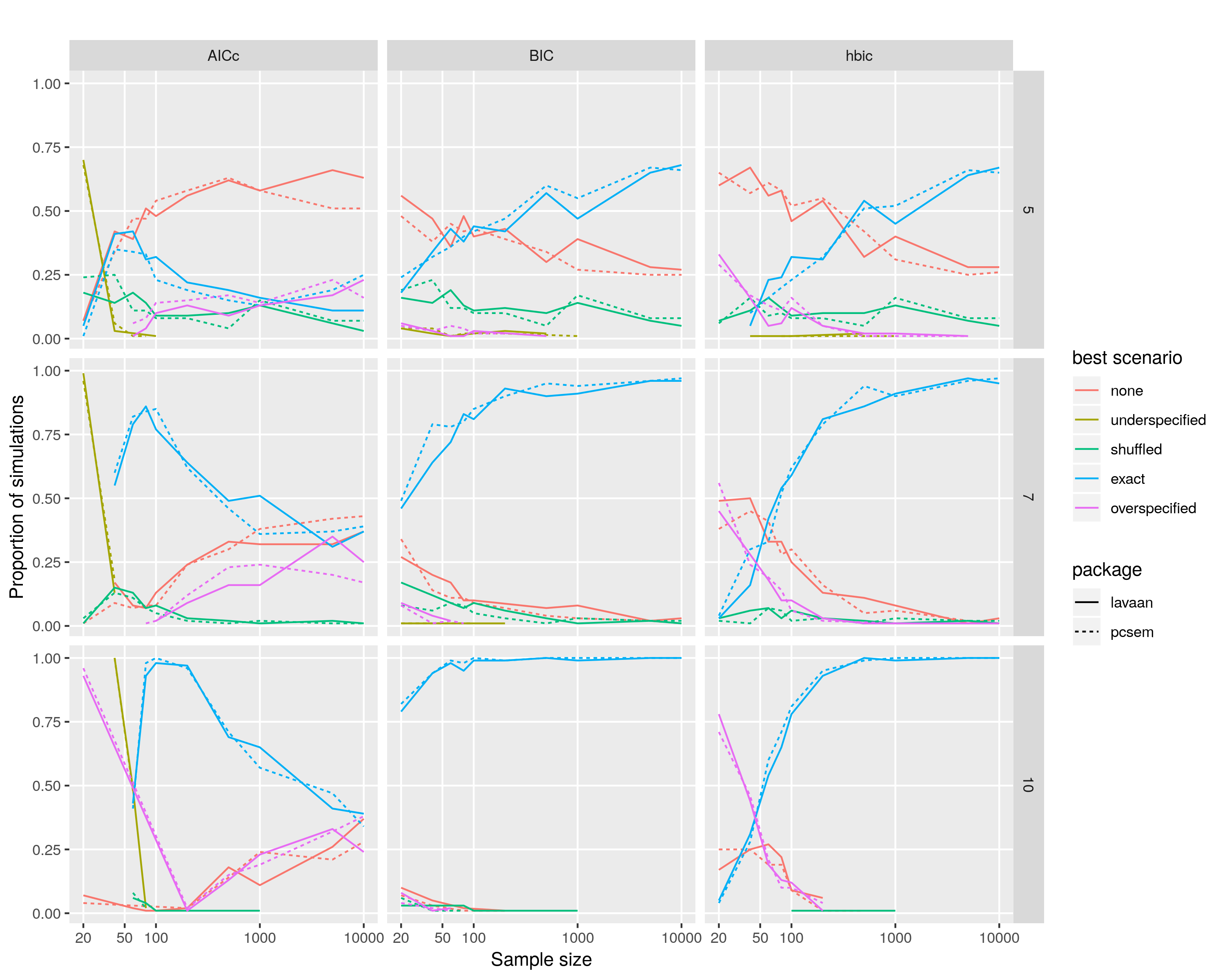}
   \caption{Effect of sample size, model complexity (5, 7 or 10 covariates in the vertical panels) and R package (lavaan or piecewiseSEM) on the proportion of simulations where the different scenarios were selected as the best scenario. Three metrics are tested (horizontal panels): AICc, BIC and HBIC.}
   \label{fig:ic}
  \end{figure}

The results for the different information criteria metrics were based on a second batch of simulations where the data were only generated once under the exact scenario and the IC values were compared across the four scenarios (excluding random) for the same data.
Exploring the proportion of simulations where each scenario was selected as the best scenario (i.e. IC value at least 2 unit below the second lowest one) revealed that, in general, AICc appears to be a poor IC metric of model selection for SEMs while BIC provided good results especially for models with larger number of covariates (Figure~\ref{fig:ic}). There were again little differences between the lavaan and the piecewiseSEM models.
In details, for low sample size (below 50) the over- or the under-specified scenario were often selected as the best scenarios with AICc. Generally for AICc the proportion of simulations where the exact scenario was selected as the best scenario showed an hump shaped pattern, the height and the position of the hump on the x and y axis depending on the number of covariates in the model. This imply that for low and large sample size AICc will not select models closer to the true data-generation process. The BIC, on the other hand, showed good results especially for models with larger number of covariates where BIC could select the exact scenario even for low sample size. BIC showed poor discriminatory capacities only for models with lower number of covariates and low sample size. The HBIC showed similar but poorer patterns than the BIC. At low sample size (below 50), the HBIC selected the exact scenario less than 50\% of the time across the different number of covariates tested. At larger sample size (especially above 100) the HBIC showed medium to good discriminatory capacities depending on the number of covariates in the model.
 
\subsection*{Signal / noise ratio effects}

Independently varying the strength of the signal and the amount of noise did not substantially affect the results. All figures relating to the signal / noise ratio variation are shown in the Appendix.  

\section*{Discussion}

A first general pattern emerging from the simulations is the broad similarity between model fitted using piecewiseSEM or lavaan. In other words, fitting SEM with piecewise assemblages of individual regressions or via the variance-covariance matrix leads to similar patterns in the explored metrics. Then, the simulations also showed the limited power of SEMs to detect effect pathways especially as model complexity increases, for instance, more than 100 of samples was needed to reach the 80\% power threshold for 10 covariates under the exact data-generation approach. This simulation showed that, for the range of sample size value used, model misspecification had much stronger impacts on model checking than sample size. Researchers should therefore acknowledge that large sample size do not protect against error in model specification. Finally, among the information criteria (IC) metrics tested, BIC provided the best capacity to select the true model and IC metrics showed better discriminatory capacities for more complex SEMs.

The first metric that researcher check on a fitted SEM is the p-value of the chi-square test on the covariance matrix for lavaan and the p-value for the Fisher's C statistic in piecewiseSEM. If this p-value is larger than 0.05 standard practice is to accept the model and explore the relationships further to unravel the relations between the covariates \citep{grace2006}. This approach is vulnerable to model misspecification. The biggest concern being that models may be readily accepted even if there is no signal in the data as revealed by the high proportion (larger than 80\% in most cases) of random models accepted. Also relying solely on one p-value to accept model fit will tend to result in building overly complex models. Indeed the simulations, showed that models more complex than the processes that generated the data will be more often accepted than models being simpler than the data-generation process. This result is to be expected in piecewiseSEM based on Fischer's C statistic that is compiling p-values of links not included in the model, so it automatically follows that this test has no safeguard against model overfitting. Finally, reversing the direction of the effects lead to large decrease in acceptance rates of the models, so model rejection might also indicate some misspecification in the directionality of the effects. In psychology, issues related with the global chi-square test and its associated p-values have been known for a long time \citep{bollen1992}. This body of literature emphasize the fact that there is no one metric that can reliably tell if the SEM fit is acceptable or not \citep[see also][]{kline1999}. These authors argued already at that time that exploration of the model component such as the magnitude of the path coefficient or the R-square should be inherent to any SEM model check. One of the most influential book on SEM in ecology \citep{shipley2000b} meticulously explores the issue of sample size and non-normality of the errors on the chi-square test and provide some solutions such as bootstrapping for complex cases. The use of such fixes is rare if not totally absent in the current ecological literature. piecewiseSEM use a different approach to model checking based on the independence claims derived from the models. I am not aware of any simulation or study exploring the impact of sample size on the Fisher's C statistic. 

The next step in SEM inference is the exploration of the fitted relations between the covariates. The result of the simulations showed that to reach a power of detecting significant paths of 80\% at least 5 data points per path coefficient should be available as a rough rule of thumb, confirming previous results \citep{fan2016}. In addition, lavaan seems to have higher power than piecewiseSEM especially at low sample size for more complex models. Interestingly, the direction of the effects had no impact on their significance. Indeed, the exact and shuffled data-generation had similar levels of significant paths. Deriving causality statements from fitted SEM should not be based on solely screening the significance of the regression coefficients. However, exploring coefficient significance is a strong test that some signal is present in the data and was captured in the models. The random data-generation revealed that the number of significant paths even under complete noise was close to the 5\% level of the type I error. So researchers should be ready to modify or reject models that, despite being accepted based on the chi-square or Fishers's C statistics, have suspiciously low number of significant paths. To my knowledge, there is little literature about the power of SEM to detect individual effects. \citet{wolf2013} reported power and bias to detect direct and indirect effects in a three latent-variables model each with three indicator variables. They reported that for detecting weak direct effects with a power of 80\% sample size should be larger than 300, for detecting the indirect effect sample size should be higher than 400. Such sample sizes are rarely reached in ecological studies that usually consist of complex models \citep{scherber2010,duffy2015}, therefore ecologists should acknowledge the power issue of SEMs and strive to gather datasets of appropriate size relative to their model complexity.

The R square metric represents the amount of variation of the data explained by the model. Ecologists readily use R-square when reporting statistical analysis, this is also true for SEM \citep{duffy2015}. The simulations showed that the average R-square was a good indicator that some signal present in the data has been extracted by the model. However, the R-square metric was largely equivalent between exact, shuffled, under- or over-specified models indicating that R-square values do not provide safeguards against model miss-specifications.

The structure of SEM imply some conditional independence between covariates, these conditional independence can be tested and provides good indication if and where a SEM fails \citep{thoemmes2017}. The simulations showed that conditional independence tests are blind to random noise, data randomly generated create no conditional independence failures. Similarly conditional independence tests will tend to favor more complex models, since models simpler than the process generating the data will be flagged by such tests, while models more complex than data-generation have very low levels of failures in conditional independence. More positively, such a metric can readily detect misspecification of the directionality of the effects. In summary, local tests of conditional independence may provide good guidance to identify missing or miss-directed links, but these tests do not prevent from model overfitting. \citet{shipley2000art} already discussed techniques to derive all conditional independence implied by direct acyclic graphs and the advantages that such an approach provides. Most interesting is the possibility to identify parts of poor fit in the model to provide guidance for further model and/or theory improvements. 

Finally, the recent surge in information theory in ecology \citep{aho2014} is also increasingly being transposed to model selection in SEMs \citep{fan2016}. The simulation results reported here show that, amongst the different information criteria (IC) metrics tested, BIC provided the best capacities to select for the true models. In general, all tested IC metrics showed better results for more complex models. BIC and HBIC capacity to detect the true models asymptotically increased with sample size, while for AICc the same relation was hump-shaped. \citet{bollen2014} compared a broader range of IC metrics and found that HBIC performed better than BIC to select the true model. The difference with the results reported here might stem from the fact that the models in \citet{bollen2014} had latent variables. In addition, \citet{bollen2014} also reported that IC metrics tended to favor overspecified models.

\subsection*{Practical recommendations}

\begin{figure}
 \includegraphics[width=\textwidth,height=.9\textheight,keepaspectratio]{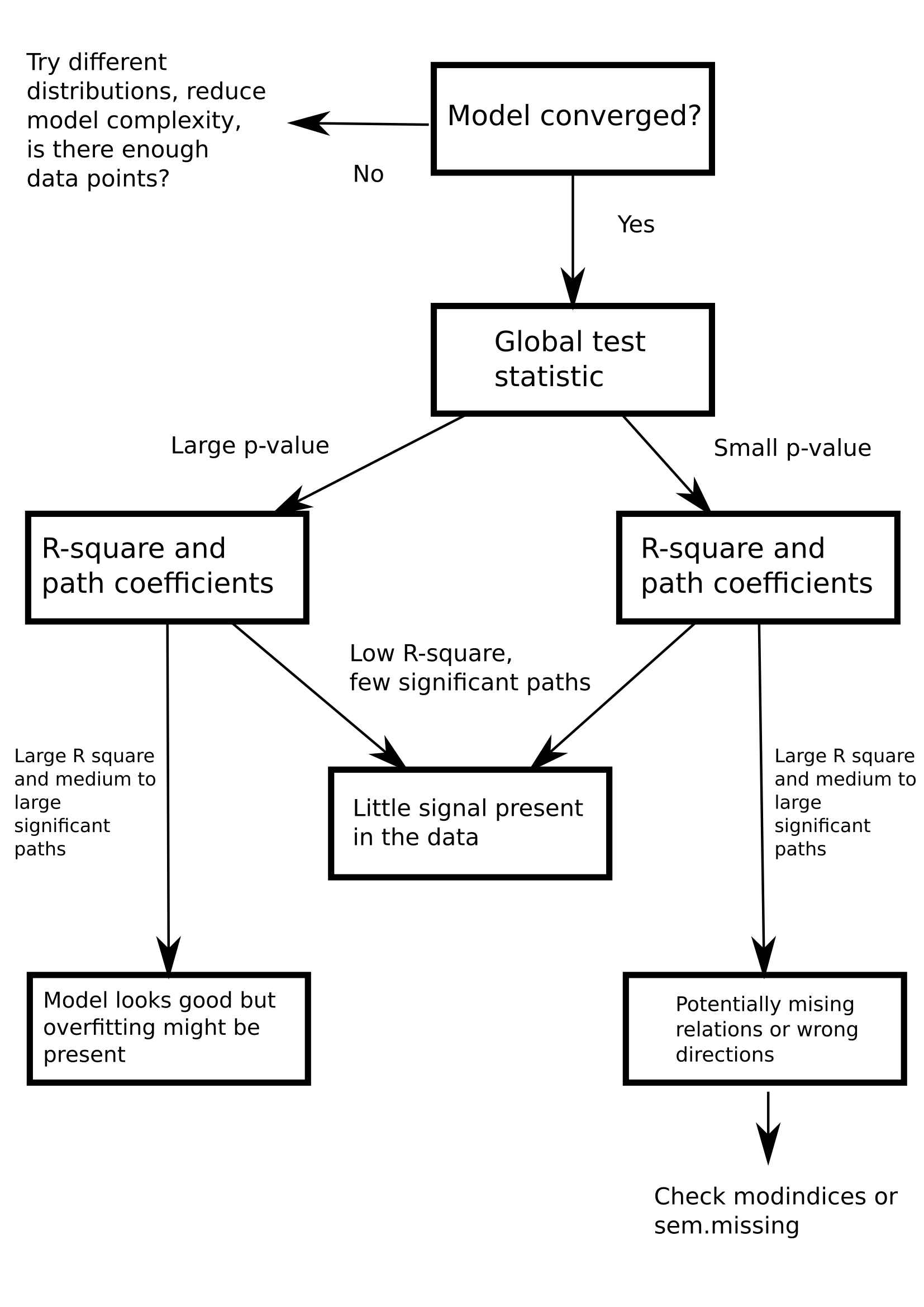}
 \caption{Flowchart describing how the informations from the different metrics studied here maybe be combined to inform about model fitness. Issues of model convergence should lead to model re-specification as the estimates of unconverged models may not be trusted.}
 \label{fig:res}
\end{figure}

Combining all results lead to some recommendations regarding post-fitting checks of structural equation models, also summarized in a flowchart (Figure~\ref{fig:res}):
\begin{itemize}
 \item The results re-emphasize the need to explore structural equation model fitness via several metrics
 \item Seemingly good fitting models with large p-values of the chi-square test or the C statistic with few or no significant paths and low R-squares imply that no real signal was extracted from the data
 \item Conversely, poorly fitting models with regards to global p-values but with a large number of significant paths, indications of un-modelled conditional dependence, and/or large R-square may indicate missing relationships between covariates (underfitting) or errors in the directionality of the effects
 \item Finally, the chi-square or Fischer's C p-value, the p-values of the paths, the R-squares and conditional independence tests do not provide safeguards against fitting models with larger numbers of relation than necessary (overfitting). Using model selection with BIC could be a solution especially for moderate to large sample size (\textgreater 100) and for complex models.

\end{itemize}

These simulation results may be of use at various stages of the data analysis. A priori, these results provide some guidance regarding the needed sample size based on expected signal / noise ratio and model complexity. But also once the model has been fitted, these results provide a mean to combine information arising from different metrics to further criticize and improve the model. 

The simulations were done under a relatively simple framework assuming only gaussian distributions with no hierarchical structure or latent variables. This allowed a direct comparison of two different model fitting techniques that would not have been possible under more complex model structure. How model misspecification affect fitness metrics in these more complex cases remain to be studied. Researchers fitting simple linear models without hierarchical structure or latent variables may use both lavaan or piecewiseSEM approaches indiscriminately. In more complex settings restrictions set by the two approaches may constrain the decision to use one or the other. For instance, latent variables structure are only available in lavaan, while piecewiseSEM allow non-normal regression and hierarchical structure.

\subsection*{Conclusion}

Structural equation modelling is a powerful inferential framework but like any other approach it does not perform magic. SEMs have different histories and traditions in different scientific fields. While SEMs have been widely used for a long time in psychology and are therefore an inherent part of the training of researcher in psychology. In other fields such as in ecology, SEMs gathered attraction more recently thanks in part to influential papers \citep{grace2016,scherber2010}, books \citep{shipley2000b,grace2006} and R packages \citep{rosseel2012,lefcheck2016}. In these fields SEMs might feel like a brave new world and little of the necessary caution is applied in deriving interpretation from the models \citep[but see][]{grace2010}. The aim of this simulation paper was to reveal the limits of routinely used metrics from fitted SEMs to model misspecification. The main conclusions are: (i) the need to combine informations from several fitness metrics to select SEMs closer to the truth, (ii) the limited power to detect effects with small sample size (\textless 100) and (iii) the need to explore local fitness of the constituting regressions in structural equation models.

\section*{Acknowledgements}

I thank Yves Rosseel and Johnathan Lefcheck for positive feedback and discussion of early simulation results. Bram Sercu, Femke Batsleer and Mike Perring for providing helpful textual corrections.

\bibliographystyle{plainnat}
\bibliography{simsem_bib.bib}

\begin{thebibliography}{26}
\providecommand{\natexlab}[1]{#1}
\providecommand{\url}[1]{\texttt{#1}}
\expandafter\ifx\csname urlstyle\endcsname\relax
  \providecommand{\doi}[1]{doi: #1}\else
  \providecommand{\doi}{doi: \begingroup \urlstyle{rm}\Url}\fi

\bibitem[Aho et~al.(2014)Aho, Derryberry, and Peterson]{aho2014}
Ken Aho, DeWayne Derryberry, and Teri Peterson.
\newblock Model selection for ecologists: the worldviews of aic and bic.
\newblock \emph{Ecology}, 95\penalty0 (3):\penalty0 631--636, 2014.
\newblock \doi{10.1890/13-1452.1}.
\newblock URL
  \url{https://esajournals.onlinelibrary.wiley.com/doi/abs/10.1890/13-1452.1}.

\bibitem[Bentler and Weeks(1980)]{bentler1980}
Peter~M Bentler and David~G Weeks.
\newblock Linear structural equations with latent variables.
\newblock \emph{Psychometrika}, 45\penalty0 (3):\penalty0 289--308, 1980.

\bibitem[Bolker(2008)]{bolker2008}
Benjamin~M Bolker.
\newblock \emph{Ecological models and data in R}.
\newblock Princeton University Press, 2008.

\bibitem[Bollen and Long(1992)]{bollen1992}
Kenneth~A Bollen and J~Scott Long.
\newblock Tests for structural equation models: introduction.
\newblock \emph{Sociological Methods \& Research}, 21\penalty0 (2):\penalty0
  123--131, 1992.

\bibitem[Bollen et~al.(2014)Bollen, Harden, Ray, and Zavisca]{bollen2014}
Kenneth~A Bollen, Jeffrey~J Harden, Surajit Ray, and Jane Zavisca.
\newblock Bic and alternative bayesian information criteria in the selection of
  structural equation models.
\newblock \emph{Structural Equation Modeling: A Multidisciplinary Journal},
  21\penalty0 (1):\penalty0 1--19, 2014.

\bibitem[Duffy et~al.(2015)Duffy, Reynolds, Bostr{\"o}m, Coyer, Cusson, Donadi,
  Douglass, Ekl{\"o}f, Engelen, Eriksson, et~al.]{duffy2015}
J~Emmett Duffy, Pamela~L Reynolds, Christoffer Bostr{\"o}m, James~A Coyer,
  Mathieu Cusson, Serena Donadi, James~G Douglass, Johan~S Ekl{\"o}f, Aschwin~H
  Engelen, Britas~Klemens Eriksson, et~al.
\newblock Biodiversity mediates top--down control in eelgrass ecosystems: a
  global comparative-experimental approach.
\newblock \emph{Ecology letters}, 18\penalty0 (7):\penalty0 696--705, 2015.

\bibitem[Eisenhauer et~al.(2015)Eisenhauer, Bowker, Grace, and
  Powell]{eisenhauer2015}
Nico Eisenhauer, Matthew~A Bowker, James~B Grace, and Jeff~R Powell.
\newblock From patterns to causal understanding: structural equation modeling
  (sem) in soil ecology.
\newblock \emph{Pedobiologia}, 58\penalty0 (2):\penalty0 65--72, 2015.

\bibitem[Fan et~al.(2016)Fan, Chen, Shirkey, John, Wu, Park, and Shao]{fan2016}
Yi~Fan, Jiquan Chen, Gabriela Shirkey, Ranjeet John, Susie~R Wu, Hogeun Park,
  and Changliang Shao.
\newblock Applications of structural equation modeling (sem) in ecological
  studies: an updated review.
\newblock \emph{Ecological Processes}, 5\penalty0 (1):\penalty0 19, 2016.

\bibitem[Grace(2006)]{grace2006}
James~B Grace.
\newblock \emph{Structural equation modeling and natural systems}.
\newblock Cambridge University Press, 2006.

\bibitem[Grace et~al.(2010)Grace, Anderson, Olff, and Scheiner]{grace2010}
James~B Grace, T~Michael Anderson, Han Olff, and Samuel~M Scheiner.
\newblock On the specification of structural equation models for ecological
  systems.
\newblock \emph{Ecological Monographs}, 80\penalty0 (1):\penalty0 67--87, 2010.

\bibitem[Grace et~al.(2016)Grace, Anderson, Seabloom, Borer, Adler, Harpole,
  Hautier, Hillebrand, Lind, P{\"a}rtel, et~al.]{grace2016}
James~B Grace, T~Michael Anderson, Eric~W Seabloom, Elizabeth~T Borer, Peter~B
  Adler, W~Stanley Harpole, Yann Hautier, Helmut Hillebrand, Eric~M Lind,
  Meelis P{\"a}rtel, et~al.
\newblock Integrative modelling reveals mechanisms linking productivity and
  plant species richness.
\newblock \emph{Nature}, 529\penalty0 (7586):\penalty0 390, 2016.

\bibitem[Hertzog(2018)]{data}
Lionel~R Hertzog.
\newblock Data from: How robust are structural equation models to model
  misspecification: a simulation study.
\newblock \emph{Zenodo}, 2018.
\newblock URL \url{http://doi.org/10.5281/zenodo.1465616}.

\bibitem[Kline and Santor(1999)]{kline1999}
Rex~B Kline and Darcy~A Santor.
\newblock Principles \& practice of structural equation modelling.
\newblock \emph{Canadian Psychology}, 40\penalty0 (4):\penalty0 381, 1999.

\bibitem[Lefcheck(2016)]{lefcheck2016}
Jonathan~S Lefcheck.
\newblock piecewisesem: piecewise structural equation modelling in r for
  ecology, evolution, and systematics.
\newblock \emph{Methods in Ecology and Evolution}, 7\penalty0 (5):\penalty0
  573--579, 2016.

\bibitem[Lin et~al.(2017)Lin, Huang, and Weng]{lin2017}
Li-Chung Lin, Po-Hsien Huang, and Li-Jen Weng.
\newblock Selecting path models in sem: A comparison of model selection
  criteria.
\newblock \emph{Structural Equation Modeling: A Multidisciplinary Journal},
  24\penalty0 (6):\penalty0 855--869, 2017.

\bibitem[Pearl(2012)]{pearl2012}
Judea Pearl.
\newblock The causal foundations of structural equation modeling.
\newblock Technical report, CALIFORNIA UNIV LOS ANGELES DEPT OF COMPUTER
  SCIENCE, 2012.

\bibitem[Petraitis et~al.(1996)Petraitis, Dunham, and
  Niewiarowski]{petraitis1996}
PS~Petraitis, AE~Dunham, and PH~Niewiarowski.
\newblock Inferring multiple causality: the limitations of path analysis.
\newblock \emph{Functional ecology}, pages 421--431, 1996.

\bibitem[Pornprasertmanit et~al.(2016)Pornprasertmanit, Miller, and
  Schoemann]{simsem}
Sunthud Pornprasertmanit, Patrick Miller, and Alexander Schoemann.
\newblock \emph{simsem: SIMulated Structural Equation Modeling}, 2016.
\newblock URL \url{https://CRAN.R-project.org/package=simsem}.
\newblock R package version 0.5-13.

\bibitem[Power(1972)]{power1972}
Dennis~M Power.
\newblock Numbers of bird species on the california islands.
\newblock \emph{Evolution}, 26\penalty0 (3):\penalty0 451--463, 1972.

\bibitem[Rosseel(2012)]{rosseel2012}
Yves Rosseel.
\newblock Lavaan: An r package for structural equation modeling and more.
  version 0.5--12 (beta).
\newblock \emph{Journal of statistical software}, 48\penalty0 (2):\penalty0
  1--36, 2012.

\bibitem[Scherber et~al.(2010)Scherber, Eisenhauer, Weisser, Schmid, Voigt,
  Fischer, Schulze, Roscher, Weigelt, Allan, et~al.]{scherber2010}
Christoph Scherber, Nico Eisenhauer, Wolfgang~W Weisser, Bernhard Schmid,
  Winfried Voigt, Markus Fischer, Ernst-Detlef Schulze, Christiane Roscher,
  Alexandra Weigelt, Eric Allan, et~al.
\newblock Bottom-up effects of plant diversity on multitrophic interactions in
  a biodiversity experiment.
\newblock \emph{Nature}, 468\penalty0 (7323):\penalty0 553, 2010.

\bibitem[Shipley(2000{\natexlab{a}})]{shipley2000art}
Bill Shipley.
\newblock A new inferential test for path models based on directed acyclic
  graphs.
\newblock \emph{Structural Equation Modeling}, 7\penalty0 (2):\penalty0
  206--218, 2000{\natexlab{a}}.

\bibitem[Shipley(2000{\natexlab{b}})]{shipley2000b}
Bill Shipley.
\newblock \emph{Cause and correlation in biology}.
\newblock Cambridge University Press, 2000{\natexlab{b}}.

\bibitem[Textor and {van der Zander}(2016)]{dagitty}
Johannes Textor and Benito {van der Zander}.
\newblock \emph{dagitty: Graphical Analysis of Structural Causal Models}, 2016.
\newblock URL \url{https://CRAN.R-project.org/package=dagitty}.
\newblock R package version 0.2-2.

\bibitem[Thoemmes et~al.(2017)Thoemmes, Rosseel, and Textor]{thoemmes2017}
Felix Thoemmes, Yves Rosseel, and Johannes Textor.
\newblock Local fit evaluation of structural equation models using graphical
  criteria.
\newblock 2017.

\bibitem[Wolf et~al.(2013)Wolf, Harrington, Clark, and Miller]{wolf2013}
Erika~J Wolf, Kelly~M Harrington, Shaunna~L Clark, and Mark~W Miller.
\newblock Sample size requirements for structural equation models: An
  evaluation of power, bias, and solution propriety.
\newblock \emph{Educational and psychological measurement}, 73\penalty0
  (6):\penalty0 913--934, 2013.

\end{thebibliography}

\appendix

\renewcommand\thefigure{A.\arabic{figure}} 
\setcounter{figure}{0}    

\section*{Appendix}

Below are reported the results for varying the signal / noise ratio by changing the amount of variation in the path coefficients (Sd\_eff) and by changning the amount of residual variation (Sd\_res). Do note that results from the main figures are based on an Sd\_eff of 2.5 and Sd\_res of 1.

\begin{figure}
 \includegraphics[width=\textwidth,keepaspectratio]{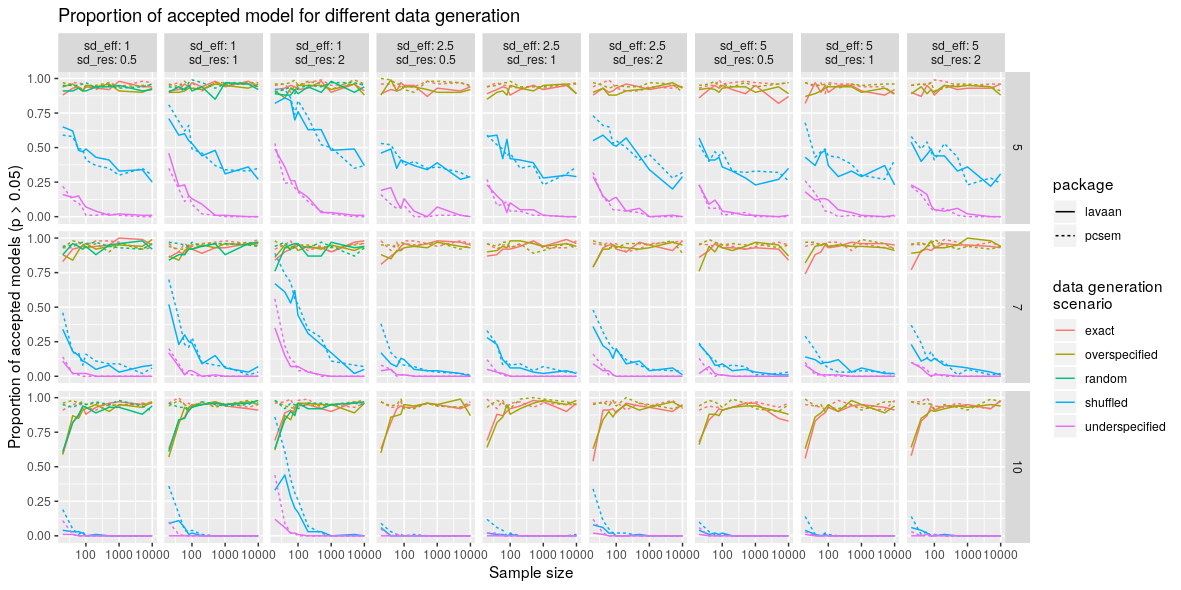}
 \caption{Proportion of accepted models with varying sample size (x-axis), data-generation (colors), package type (linetype), number of covariates (rows) and signal / noise ratio (columns).}
\end{figure}

\begin{figure}
 \includegraphics[width=\textwidth,keepaspectratio]{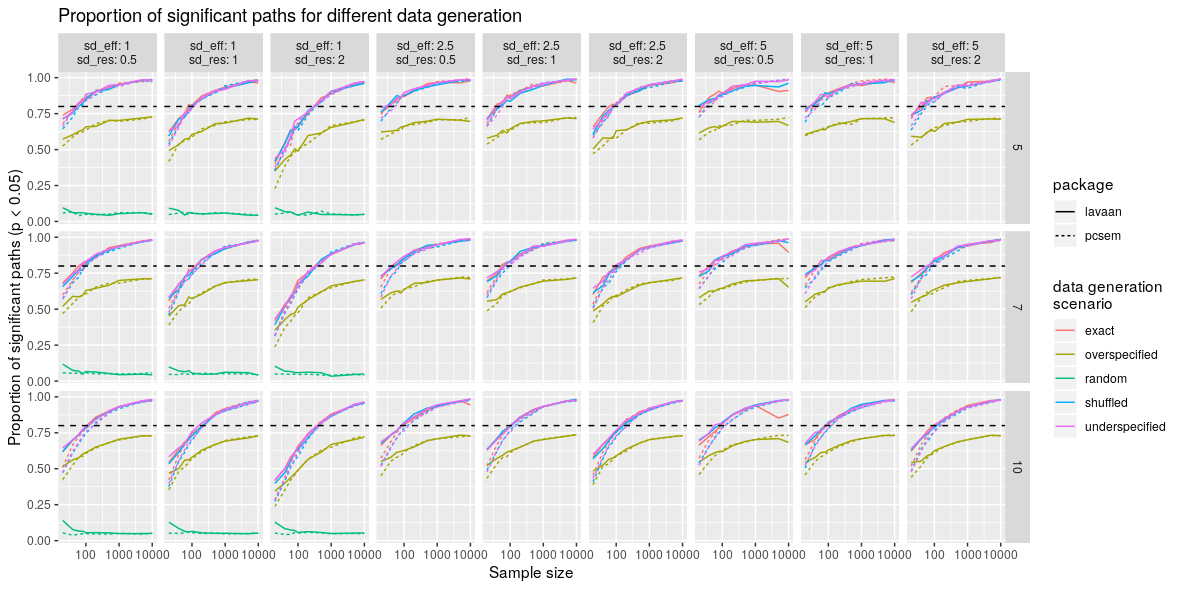}
 \caption{Proportion of accepted model paths with varying sample size (x-axis), data-generation (colors), package type (linetype), number of covariates (rows) and signal / noise ratio (columns).}
\end{figure}

\begin{figure}
 \includegraphics[width=\textwidth,keepaspectratio]{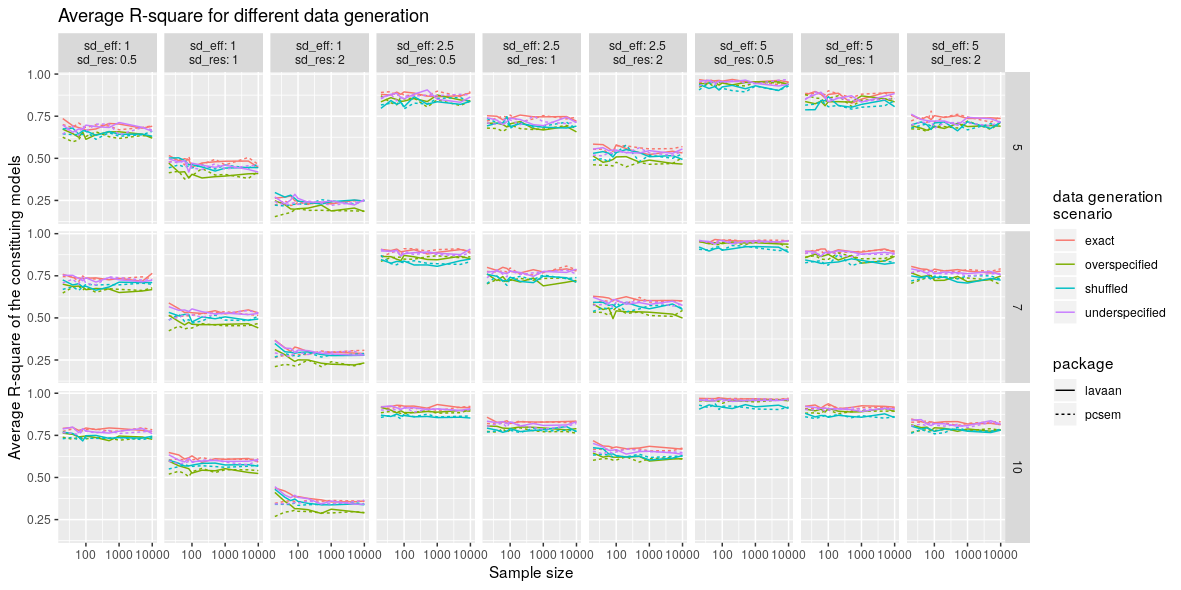}
 \caption{Average R-square values with varying sample size (x-axis), data-generation (colors), package type (linetype), number of covariates (rows) and signal / noise ratio (columns).}
\end{figure}

\begin{figure}
 \includegraphics[width=\textwidth,keepaspectratio]{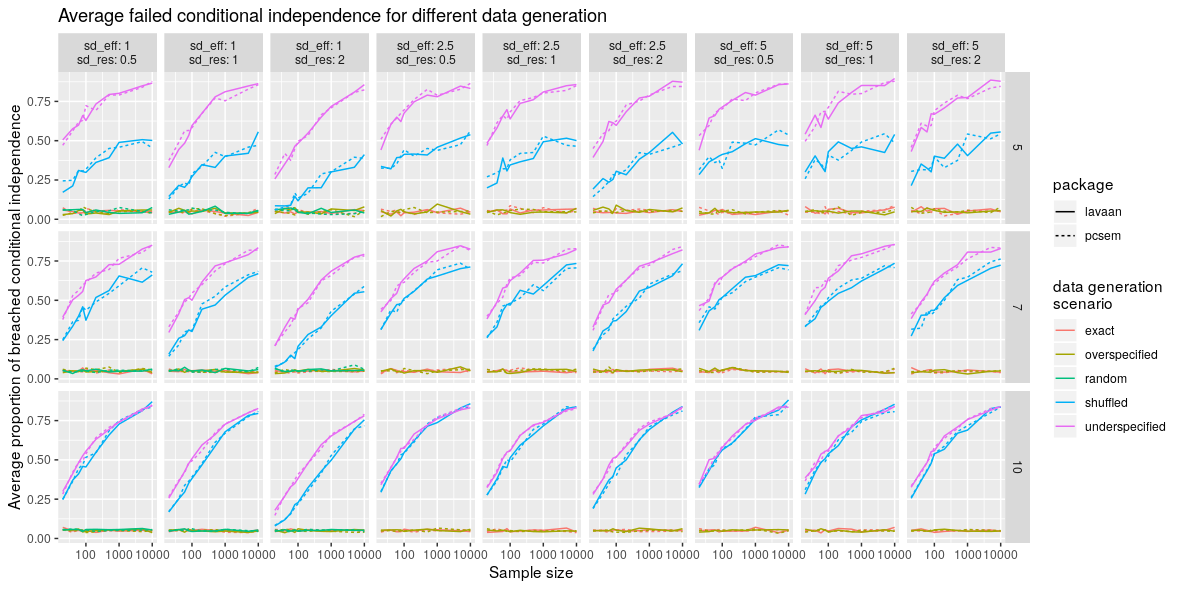}
 \caption{Proportion of failed conditional independence tests with varying sample size (x-axis), data-generation (colors), package type (linetype), number of covariates (rows) and signal / noise ratio (columns).}
\end{figure}

\begin{figure}
 \includegraphics[width=\textwidth,keepaspectratio]{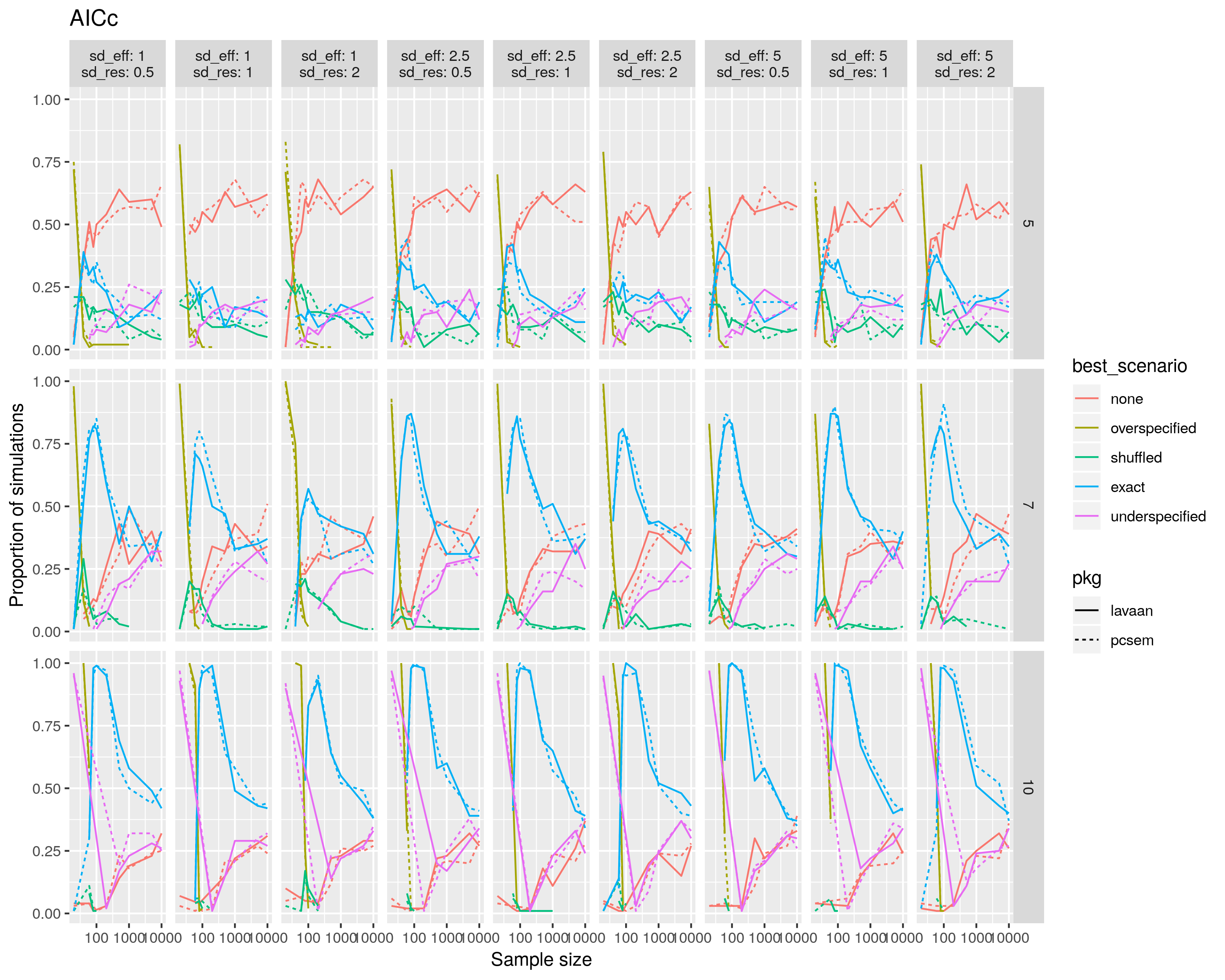}
 \caption{Proportion of simulations when the different scenarios (colors) were selected as the best scenario based on AICc with varying sample size (x-axis), package type (linetype), number of covariates (rows) and signal / noise ratio (columns).}
\end{figure}

\begin{figure}
 \includegraphics[width=\textwidth,keepaspectratio]{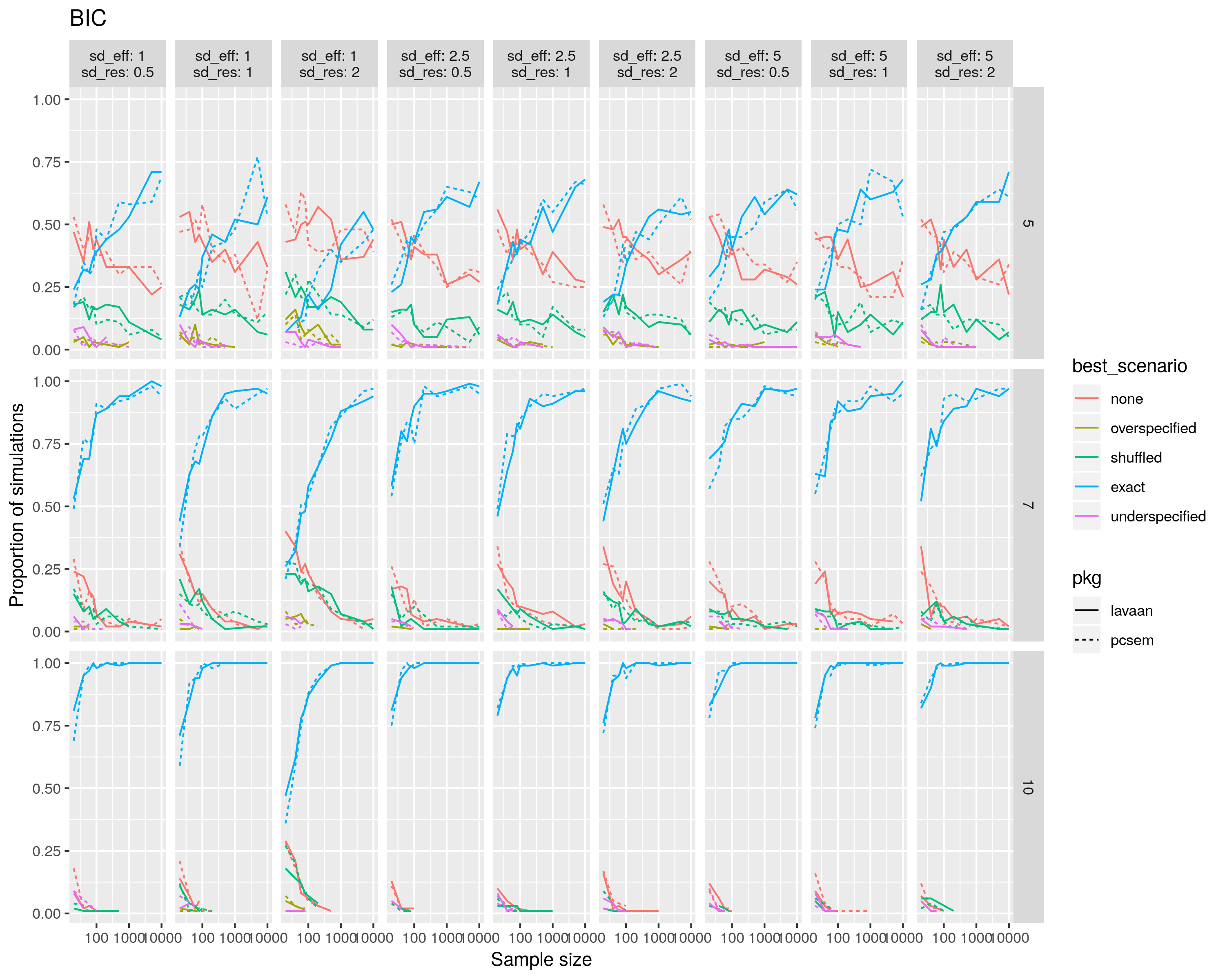}
 \caption{Proportion of simulations when the different scenarios (colors) were selected as the best scenario based on BIC with varying sample size (x-axis), package type (linetype), number of covariates (rows) and signal / noise ratio (columns).}
\end{figure}

\begin{figure}
 \includegraphics[width=\textwidth,keepaspectratio]{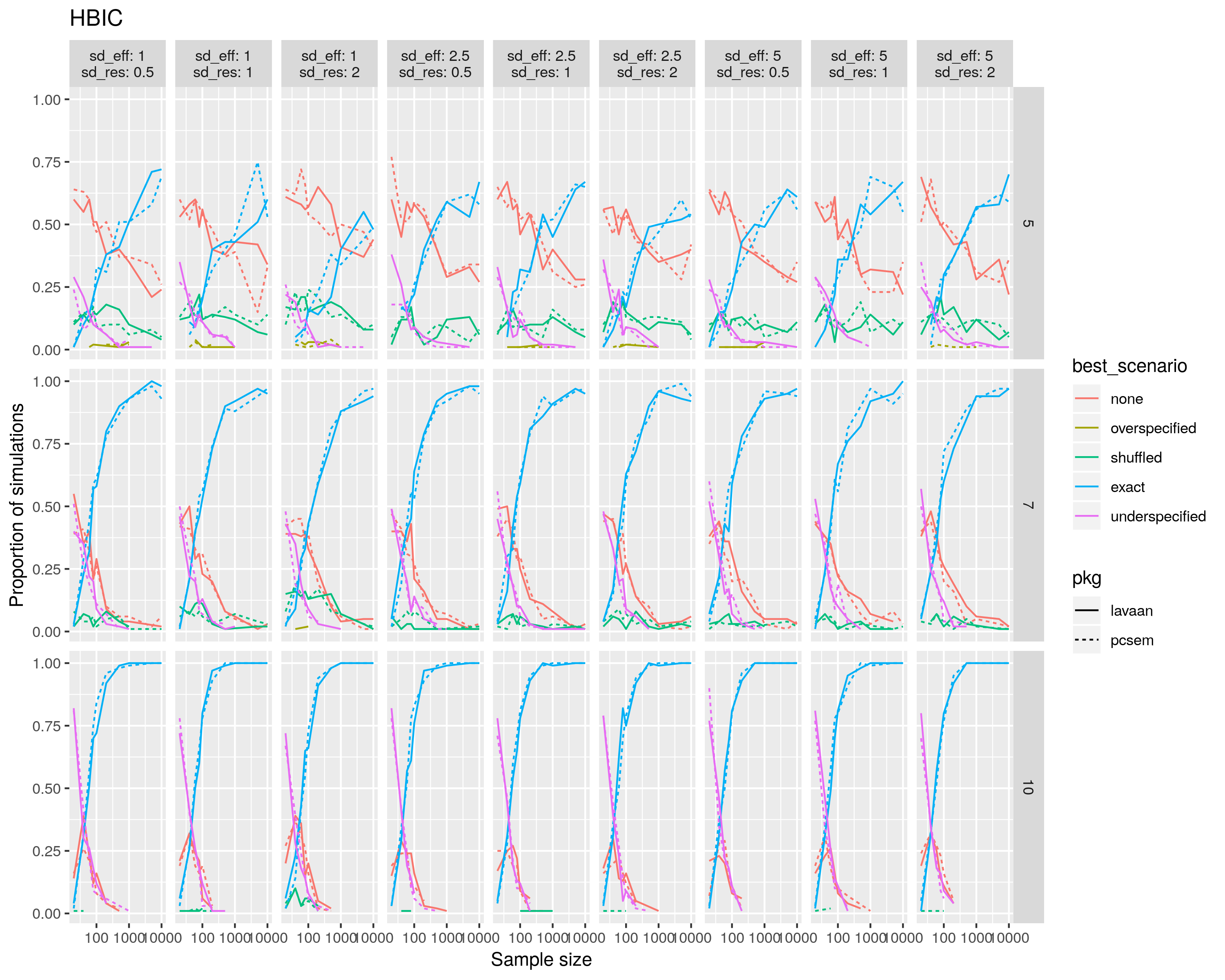}
 \caption{Proportion of simulations when the different scenarios (colors) were selected as the best scenario based on HBIC with varying sample size (x-axis), package type (linetype), number of covariates (rows) and signal / noise ratio (columns).}
\end{figure}

\end{document}